\documentstyle[aps,graphicx,prb,floats]{revtex}

\begin{document}
\draft

\twocolumn[\hsize\textwidth\columnwidth\hsize\csname
@twocolumnfalse\endcsname

\title{First-principles study of (BiScO$_3$)$_{1-x}$-(PbTiO$_3$)$_x$
piezoelectric alloys}

\author{Jorge \'I\~niguez,$^1$ David Vanderbilt,$^1$ and
L. Bellaiche$^2$}

\address{$^1$Department of Physics and Astronomy, Rutgers University,
Piscataway, New Jersey 08854-8019, USA\\ $^2$Physics Department,
University of Arkansas, Fayetteville, Arkansas 72701, USA}

\date{January 31, 2003}

\maketitle

\begin{abstract}
We report a first-principles study of a class of
(BiScO$_3$)$_{1-x}$-(PbTiO$_3$)$_x$ (BS-PT) alloys recently proposed by
Eitel {\it et al.}\ as promising materials for piezoelectric actuator
applications. We show that (i) BS-PT displays very large structural
distortions and polarizations at the morphotropic phase boundary (MPB)
(we obtain a $c/a$ of $\sim$1.05-1.08 and $P_{\rm tet}\approx
0.9$~C/m$^2$); (ii) the ferroelectric and piezoelectric properties of
BS-PT are 
dominated by the onset of hybridization between Bi/Pb-6$p$ and O-2$p$
orbitals, a mechanism that is enhanced upon substitution of Pb by Bi;
and (iii) the piezoelectric responses of BS-PT and 
Pb(Zr$_{1-x}$Ti$_x$)O$_3$ (PZT) at the MPB are
comparable, at least as far as the computed values of the
piezoelectric coefficient $d_{15}$ are concerned.  While our results
are generally consistent with experiment, they also suggest that
certain intrinsic properties of BS-PT may be even better than has
been indicated by experiments to date.  We also discuss results for
PZT that demonstrate the prominent role played by Pb displacements in
its piezoelectric properties.
\end{abstract}


\vskip2pc]

\narrowtext

\marginparwidth 2.7in
\marginparsep 0.5in

\section{Introduction}

Perovskite alloys based on PbTiO$_3$ (PT) are of considerable interest
for applications as piezoelectric actuator
materials.\cite{uch96,par02} The phase diagrams of the most
technologically important alloys are characterized by a morphotropic
phase boundary (MPB) separating the PT-rich tetragonal phase, in which
the polarization lies along a $\langle001\rangle$ direction, from the
PT-poor rhombohedral phase, in which the polarization is along a
$\langle111\rangle$ direction.\cite{noh02} Because the structural
transition between these two phases brings about a large
electromechanical response, materials with a composition lying close
to the MPB are the preferred ones for applications.  Examples of such
materials are PZT or PZN-PT, in which PT is alloyed with PbZrO$_3$
(PZ) or Pb(Zn$_{1/3}$Nb$_{2/3}$)O$_3$, respectively.

Recently, attention has been drawn to a new class of materials in
which PT is alloyed with Bi-based perovskites, the best-studied
example being (BiScO$_3$)$_{1-x}$-(PbTiO$_3$)$_x$ (BS-PT).\cite{eitel}
The appeal of BS-PT is twofold: (i) it appears to 
have piezoelectric properties comparable to those of PZT and PZN-PT
in quality, and (ii) its dielectric and piezoelectric properties
should be more robust, over a wider temperature range, than those
of PZT and PZN-PT.  The physical reason behind (ii) is that the
Curie temperature within the MPB composition range ($T^{\rm
MPB}_C$) is higher in BS-PT ($\sim$~450$^\circ$C) than in PZT
($\sim$~400$^\circ$C) or PZN-PT ($\sim$~200$^\circ$C), suggesting
that BS-PT at room temperature will age more slowly, its properties
will be more temperature-independent, etc.\cite{par02,eitel}
A possible drawback is that BS-PT is evidently not thermodynamically
stable in the perovskite crystal structure over the entire range of
composition.  However, for applications purposes it is only
important that the perovskite structure can be obtained for
compositions near the MPB, and this is indeed the case.

The authors of Ref.~\onlinecite{eitel} synthesized
BS-PT guided by an empirical crystal-chemistry rule suggesting that
since Bi$^{3+}$ is too small an ion to form a cubic perovskite with
Sc$^{3+}$, it will have a strong tendency to move off-center from its
high-symmetry position, yielding a high structural transition
temperature. While perhaps overly simplistic, this reasoning is
undoubtedly partially correct and is likely to be useful for materials
engineering. However, it is unclear how to explain the large
piezoelectric responses found in BS-PT based on ion-size
considerations alone. The ferroelectric properties of materials like
PZT are known to be related to the partial covalency of some
bonds,\cite{coh92,pos94} and the situation in BS-PT should be
similar. On the other hand, most of the relevant piezoelectric alloys
have only Pb on the A site of their perovskite structure, and the
chemistry of Pb is known to be very important to their
properties.\cite{ega02} The substitution of Pb by Bi in BS-PT is thus
of particular interest in view of the good piezoelectric properties of
this material.

We present here a first-principles study of BS-PT. We find that Bi
plays a crucial role, and in particular that hybridization between
Bi-6$p$ and O-2$p$ orbitals is the driving force for the ferroelectric
instabilities of BS-PT, allowing for very large polarizations and
responses. Our results are generally consistent with experiment, but
they also suggest that the intrinsic ferroelectric and piezoelectric
properties of BS-PT alloys might be even better than those measured
experimentally to date.

\begin{table*}
\caption{Relaxed structure and polarization for BiScO$_3$ (BS),
PbTiO$_3$ (PT), PbZrO$_3$ (PZ), BiYO$_3$ (BY), a VCA BS-PT alloy at
$x=0.5$ (BS-PT (VCA)), a 10-atom BS-PT supercell (BS-PT (SC)),
a VCA PZT alloy at $x=0.5$ (PZT (VCA)), and a 10-atom PZT
supercell (PZT (SC)).  (See text for details.)  Reported are the
tolerance factor $t$; the energy difference $\Delta E$ (in eV)
per 5-atom cell
relative to the cubic phase, and the polarization $P$ (in
C/m$^{2}$), for tetragonal and rhombohedral phases; and the $c/a$
ratio and the rhombohedral angle $\alpha$ (in degrees) for the
tetragonal and rhombohedral phases respectively. BS-PT (SC)
has two inequivalent rhombohedral phases; the reported results
correspond to the lower-energy phase.}
\vskip 1mm
\begin{tabular}{ld|ddd|ddd}
& & \multicolumn{3}{c|}{Tetragonal phase}
  & \multicolumn{3}{c}{Rhombohedral phase} \\
System & $t$ & $\Delta E$ & $c/a$ & $P$ & $\Delta E$ & $\alpha$ & $P$\\
\hline
BS & 0.907 & $-$1.124 & 1.285 & 0.93 & $-$1.353 & 88.6 & 0.47\\
PT & 1.027 & $-$0.060 & 1.049 & 0.82 & $-$0.047 & 89.6 & 0.72\\
PZ & 0.970 & $-$0.214 & 1.035 & 0.73 & $-$0.274 & 89.5 & 0.78\\
BY & 0.845 & $-$2.260 & 1.376 & 0.95 & $-$2.896 & 88.7 & 0.49\\
BS$-$PT (VCA)& & $-$0.376 & 1.114 & 0.92 & $-$0.412 & 89.1 & 0.60\\
BS$-$PT (SC) & & $-$0.472 & 1.079 & 1.06 & $-$0.565 & 89.4 & 1.05\\
PZT (VCA)  & & $-$0.060 & 1.026 & 0.69 & $-$0.061 & 89.6 & 0.67\\
PZT (SC)   & & $-$0.096 & 1.032 & 0.73 & $-$0.104 & 89.5 & 0.72
\end{tabular}
\label{tab1}
\end{table*}

We present our results in the form of a systematic comparison of the
properties of BS-PT with those of PZT.  Section~II describes the
first-principles methods employed. In Sec.~III we study the
ferroelectric instabilities of pure BS, PT, and PZ, introducing some
corresponding calculations on BiYO$_3$ (BY) as an aid to the
discussion. In Sec.~IV we discuss the ferroelectric instabilities of
the alloys and make contact with the experimental results, while
Sec.~V is devoted to the piezoelectric properties of the alloys near
the MPB. We summarize and conclude in Sec.~VI.

\section{Methods}

The calculations were performed within a plane-wave implementation of
the local-density approximation (LDA) to density-functional theory,
using ultrasoft pseudopotentials~\cite{van90} to represent the ionic
cores. The following electronic states were included in the
calculation: the 2$s$ and 2$p$ states of O, the 3$s$, 3$p$, 3$d$, and
4$s$ states of Sc and Ti, the 4$s$, 4$p$, 4$d$, and 5$s$ states of Y
and Zr, and the 5$d$, 6$s$, and 6$p$ states of Pb and Bi.
For the 5-atom unit cell calculations, we used a 35~Ry plane-wave
cutoff and a $6\times 6\times 6$ k-point grid for Brillouin zone
integrations. We checked that these choices yield converged
results. We also made some calculations with a rocksalt-ordered (fcc)
10-atom unit cell, for which we used a 30~Ry cutoff and a $4\times
4\times 4$ k-point grid.
Atomic relaxations were considered to be converged when the residual
forces were smaller than $5\times 10^{-5}$~a.u. Polarizations were
computed using the Berry-phase expression of King-Smith and
Vanderbilt~\cite{kin93} and always using a dense-enough k-point mesh
along the direction of the strings.  The character of the electron
energy bands was estimated by computing the partial density associated
with the pseudopotential projectors of specific atoms (see Eq.~(19) of
Ref.~\onlinecite{van90} for guidance).

Calculations using the virtual-crystal approximation (VCA) were
implemented as in Ref.~\onlinecite{bel00a}. In the VCA one works with
virtual atoms defined by a pseudopotential that is a weighted average
of those of real atoms.  Such a procedure may lead to meaningful
estimates of the average properties of an alloy provided that the
valence orbitals of the real atoms forming the virtual atom are
similar in character. This can be expected to be true for the pairs
Bi-Pb and Sc-Ti in BS-PT. Also, both groups BiSc and PbTi have a total
nominal ionic charge of $+6$, allowing them to be mixed in arbitrary
proportions. These requirements are also satisfied by PZT, which has
previously been studied within the VCA.\cite{bel00a,ram00} It is
important to note that the VCA is not expected to render an accurate
description of the electronic structure of a truly disordered alloy,
but rather to provide us with qualitative energetic and structural
trends. Also, the VCA is expected to work better for isovalent pairs of
atoms (e.g., Zr-Ti) than for heterovalent pairs (e.g., Bi-Pb or
Sc-Ti), which makes BS-PT a particularly challenging system. For this
reason, we assess the reliability of our VCA results by comparing them
with some supercell calculations and available experimental
information.

\section{Pure compounds}

As a preliminary step towards understanding BS-PT, we first consider
pure BS (in the perovskite structure\cite{fn:pureBS}) and compare
with corresponding calculations for pure PT and PZ.  The first three
lines of Table~\ref{tab1} show the structural and polarization results
obtained by relaxing these systems within ferroelectric phases of
tetragonal ($P4mm$) or rhombohedral ($R3m$) symmetry, always
maintaining a 5-atom unit cell. The atomic relaxations corresponding
to the tetragonal phases of BS and PT are given in Table~\ref{tab2}.

\begin{table}[b!]
\caption{Lattice constant $c$ and atomic coordinates (relative to
A site, in units of $c$) in the tetragonal phase of several
systems (following the notation of Table~\protect\ref{tab1}). Note
that the ideal cubic-perovskite coordinates are $z$(B)=0.5, $z$(O$_x$,
O$_y$)=0.5, and $z$(O$_z$)=0.0.}
\vskip 1mm
\begin{tabular}{lddddd}
       &    &    &    & (VCA) & (VCA)\\
       & BS & PT & BY & BS-PT & PZT \\
\hline
$c$ (a.u.) & 9.14 & 7.65 & 10.00 & 8.07 & 7.65\\
$z$(B) & 0.573 & 0.535 & 0.575 & 0.571 & 0.547\\
$z$(O$_x$, O$_y$) & 0.729 & 0.606 & 0.761 & 0.682 & 0.610\\
$z$(O$_z$) & 0.177 & 0.094 & 0.180 & 0.145 & 0.085
\end{tabular}
\label{tab2}
\end{table}

For BS, we find the rhombohedral phase to be lower in energy than
the tetragonal phase, which agrees with the experimental fact that
BS-rich BS-PT is rhombohedral, and we find
that the relaxations are very large.  This is reflected both in the
structural data (we obtain $c/a=1.29$ for the tetragonal phase of
BS, compared with 1.05 for PT; see also the atomic displacements in
Table~\ref{tab2}) and in the energetics ($\Delta E_{\rm tet}$ and
$\Delta E_{\rm rho}$ of BS are one order of magnitude larger than
for PT and PZ).
These results suggest that the cubic perovskite structure is not a
very natural one for BS, a conclusion which is reinforced by
experimental indications that pure BS at zero pressure is not even
thermodynamically stable.\cite{eitel}

The polarization associated with the tetragonal phase of BS is
large,\cite{fn:quantum} while that of the rhombohedral phase is
relatively small. In both cases, the largest contribution to $P$ comes
from the displacement of Bi relative to its neighboring O ions (O$_x$
and O$_y$ in Table~\ref{tab2}). More specifically, in the cubic phase
of BS all effective charges are normal, i.e., with values close to the
nominal ionic charges, except for those of Bi (6.4) and O$_x$ and
O$_y$ ($-$3.2 along the polarization direction). Moreover, these are
the ions with the largest relative displacements associated with the
ferroelectric instability (see Table~\ref{tab2}). The differnce in $P$
between the two phases reflects a significantly smaller charge
transfer when the Bi ion approaches three O ions (rhombohedral phase)
instead of four (tetragonal phase).

\begin{figure}
\begin{center}
\includegraphics[width=3.3in]{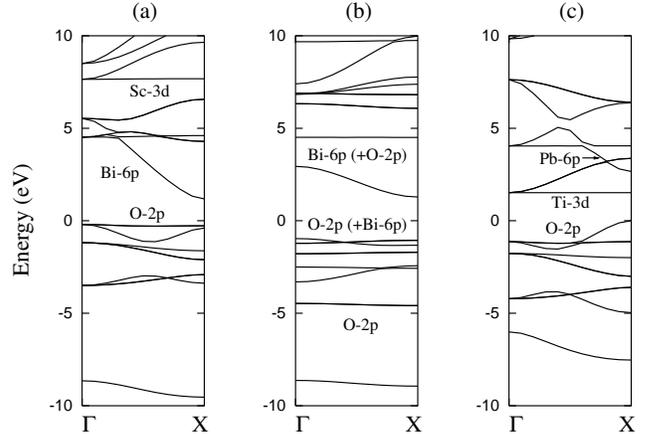}
\end{center}
\vskip 1mm
\caption{Near-gap electronic band structure along the
$\Gamma$-X direction for (a) cubic BS, (b) tetragonal BS, and (c)
cubic PT.  Zero of energy is at the valence-band maximum.}
\label{fig1}
\end{figure}

The fundamental role of Bi in the ferroelectric instabilities of BS is
best appreciated in the electronic band-structure data shown in
Fig.~\ref{fig1}. As in PT (panel~c), the top valence states of the
cubic phase of BS (panel~a) are O-2$p$ in character. However, the
lowest unoccupied states of BS are Bi-6$p$ like, at variance with the
Ti-3$d$ bands we find in PT. This fact, together with the relatively
small gap of cubic BS, results in ferroelectric instabilities mainly
characterized by a hybridization between Bi-6$p$ and O-2$p$ orbitals,
not the one involving $d$ bands of the B cation that is typical of
most other ferroelectric perovskites.\cite{pos94,kin94,hil99} In the
tetragonal phase of BS (panel~b), the gap opens considerably and the
top valence bands increase their Bi-6$p$ character. The strong
Bi-6$p$--O-2$p$ hybridization is also reflected in the fact that the
effective charge of Bi in the tetragonal phase recovers a normal value
(2.9 along the polarization direction).

As indicated earlier, Eitel and co-workers chose to study BS-PT
because of the large size mismatch between Bi and Sc.\cite{eitel} Such
a mismatch is conveniently quantified in terms of the tolerance factor
\begin{equation}
t = \frac{R_{\rm A} + R_{\rm O}}{\sqrt{2}(R_{\rm B}+R_{\rm O})},
\end{equation}
where $R_{\rm O}$, $R_{\rm A}$ and $R_{\rm B}$ are the ionic radii of
oxygen and of the A and B cations respectively.\cite{sha76}  That is,
$t$ is the ratio between the ideal cubic lattice parameters based on
A--O vs.~B--O bonding alone, with $t$=1 corresponding to high
perovskite stability. As can be seen from Table~\ref{tab1}, the
relatively low value of $t$=0.907 for BS implies that the Bi ion should
be very unstable in the high-symmetry position of the cubic perovskite
structure, consistent with our findings.  We also studied BiYO$_3$
(BY), for which $t$=0.845, and found even stronger ferroelectric
instabilities that are very similar in character to those of BS (see
Tables~\ref{tab1} and~\ref{tab2}). However, $P_{\rm tet}$ and $P_{\rm
rho}$ of BY are almost identical to those of BS, not larger. Also, the
distance between Bi and its bonded O neighbors is roughly the same in
the ferroelectric phases of BS and BY (4.33 and 4.35~a.u.,
respectively, in the tetragonal phase). These two facts seem to
indicate that the onset of the Bi-$6p$--O-$2p$ hybridization (which
also dominates the development of polarization in BY) is not very
$t$-dependent, suggesting that the large polarizations of BS and BY
cannot be attributed mainly to a size-mismatch effect.  Instead, they
are strongly linked to the particular chemistry of Bi, i.e., its strong
tendency to bond covalently with the surrounding oxygen anions.

\section{Alloys: Ferroelectric instabilities}

We now turn to a study of the BS-PT alloy within the virtual crystal
approximation (VCA).\cite{bel00a,ram00}  We also carry out parallel
VCA calculations on the PZT alloy as a relevant reference.

\begin{figure}
\begin{center}
\includegraphics[width=3.1in]{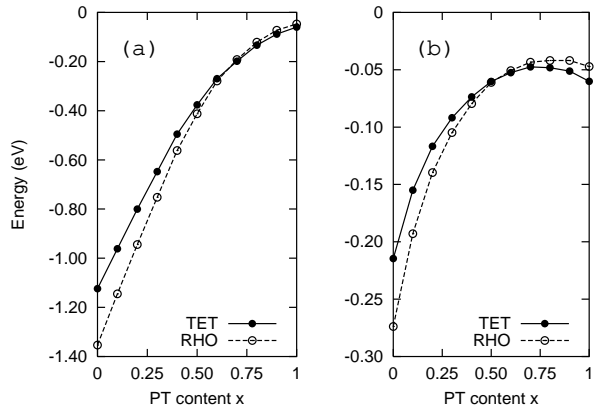}
\end{center}
\vskip 1mm
\caption{Calculated VCA energies of the tetragonal and rhombohedral
phases, relative to the cubic phase, as a function of PT content for
BS-PT (panel~a) and PZT (panel~b). Note the two different energy
scales.}
\label{fig2}
\end{figure}

Figure~\ref{fig2} shows the energies of the tetragonal and rhombohedral
ferroelectric phases of BS-PT and PZT. (The cubic phase, which is
material- and composition-dependent, is chosen as the zero of energy.)
Structural and polarization results corresponding to those obtained for
the pure compounds are shown in Tables~\ref{tab1} and~\ref{tab2}.

Associating the MPB composition with the crossing of the energy curves
for the two ferroelectric phases in Fig.~\ref{fig2}, we estimate
$x_{\rm MPB}\approx 0.65$ for BS-PT, while the experimental value is
0.64.\cite{eitel} Similarly, for PZT we get $x_{\rm MPB}\approx 0.55$,
to be compared with experimental values from about 0.45 to about 0.52
(the range of stability of the intermediate monoclinic phase of
PZT\cite{noh02}). Taking into account all the approximations
involved in our estimation,
we must regard the agreement in the case of BS-PT as partly
fortuitous. Nevertheless, these results suggest that our approach is
adequate.

Figure~\ref{fig2} also shows that the energy reductions associated
with the ferroelectric instabilities of BS-PT are considerably larger
than those of PZT. In particular, at the MPB we have $\Delta E^{\rm
MPB}=-$0.234~eV for BS-PT and $-$0.056~eV for PZT. These results are
reminiscent of the highly unstable cubic phase of BS previously
discussed. 

It is tempting to try to connect these numbers with the
Curie temperatures of BS-PT and PZT at the MPB. We know that, in
principle, the thermal energy required for the system to remain in the
high-symmetry paraelectric phase is of the order of $|\Delta E^{\rm
MPB}|$, which suggests that $T_C^{\rm MPB}$ of BS-PT is about 5~times
larger than that of PZT, i.e., higher than 1000$^\circ$C. However, we
also know that in systems in which there is competition between
different structural instabilities, such an argument can grossly
overestimate the transition temperature. In PZT, for example, it is
known that there is a competition between the ferroelectric
rhombohedral (FE-rho) phase and an antiferrodistortive (AFD) phase
involving rotations of the oxygen octahedra.\cite{for01} Such a
competition also exists in BS-PT, since we find that this
material presents a strong AFD instability. Taking PZT as a reference,
one could try to estimate the effect of this competition in BS-PT in
the following way. The AFD phase supposes a strain
of the high-symmetry structure that is opposite to that of the FE-rho
phase. At $x$=0.5 for BS-PT the volume per 5-atom cell of the FE-rho
phase at the MPB is 415~a.u.,
while it is 396~a.u.\ for the AFD phase, resulting in a volume
ratio of 1.05. This ratio is 1.01 in PZT, which suggests that the
FE-rho--AFD competition will be stronger in BS-PT than in
PZT. Energetically, though, we find the opposite trend. The FE-rho
phase is more favorable by 0.15~eV in the case of BS-PT, while in PZT
FE-rho is preferred to AFD by only 0.02~eV. Hence, we are unable to
determine in which of the two systems the reduction of $T_C^{\rm MPB}$
caused by the FE-rho--AFD competition should be larger. 
On the other hand, there is another mechanism missed by the VCA and
which could significantly lower the transition temperatures of
BS-PT. Because BS-PT is a disordered {\sl heterovalent} alloy,
internal electric fields will be present in it. Such fields will cause
local frustrations and enhance competition between different
polarization directions, resulting in a decrease of
$T_C$.\cite{ini01,geo01} In {\sl isovalent} PZT the internal fields
and their effects are bound to be much smaller. Quantifying these
effects would require a full statistical-mechanics study of the
disordered alloys, but that is beyond the scope of this work. In
conclusion, all we can say about the issue of the transition
temperatures is that our results indicate that $T_C^{\rm MPB}$ of
BS-PT could well be significantly larger than that of PZT.

Let us now discuss the structural and polarization data. The ionic and
strain relaxations of BS-PT are much larger than those of PZT (see
Tables~\ref{tab1} and~\ref{tab2}). Importantly, at the MPB composition
we find $c/a=$1.079 and $P_{\rm tet}=0.875$~C/m$^2$ for BS-PT, to be
compared, respectively, with 1.027 and 0.687~C/m$^2$ obtained for
PZT. The very large values calculated for BS-PT in the MPB region
suggest that its piezoelectric response could be very large too,
assuming the usual picture in which this arises from easy polarization
rotation associated with the near-degeneracy between
tetragonal and rhombohedral phases.\cite{fu00,bel00b} In fact,
according to our results, it would not be surprising to find that
BS-PT has even better piezoelectric properties than PZT.

Here we have to note that, while our results for PZT are in reasonable
agreement with experiment (at the MPB, $c/a$ and $P$ of PZT are
measured to be about 1.025 and 0.75~C/m$^2$, respectively\cite{capzt,ber59}),
our results for BS-PT significantly deviate from
experimental values. Ref.~\onlinecite{eitel} reports $c/a\approx
1.023$ and $P_{\rm tet}\approx 0.32$~C/m$^2$ at the MPB of BS-PT,
values that are much smaller than those that result from our VCA
calculations. In view of this discrepancy, we decided to asses the
reliability of the VCA by studying a 10-atom supercell of BS-PT at
50/50 composition.  We have chosen an fcc supercell in which there is
rocksalt ordering both on the A sublattice (Bi/Pb) and on the B
sublattice (Sc/Ti).  The resulting supercell has $T_d$ symmetry, so
that all tetragonal ferroelectric domain orientations remain
degenerate, while the the rhombohedral domains split in two groups of
four degenerate orientations. (Here we report results for the
lower-energy rhombohedral domains only.)
The results, summarized in Table~\ref{tab1}, display a significant
reduction in the $c/a$ of the tetragonal phase, but not large enough
to reconcile theory with experiment. The VCA error is presumably
maximum at $x=0.5$, so at $x_{\rm MPB}$ we would expect that $c/a$
should be at least 1.045, still significantly larger than the experimental
value. Regarding the polarization, the 10-atom supercell values
are actually larger than the VCA results.

In Table~\ref{tab1} we also included results corresponding to the
analogous 10-atom supercell calculations for PZT. As
expected,\cite{bel00a} the VCA is more accurate for PZT (isovalent
alloy) than for BS-PT (heterovalent). This is further confirmed by the
atomic relaxation data (not shown here).

Hence, for BS-PT a relatively large discrepancy between theory and
experiment remains.  Its origins are most probably related to the
differences between the idealized crystalline samples we consider and
the ceramic samples actually grown (the internal fields present in
real disordered samples are likely to play a particularly relevant
role). In any  case, we hope that the very 
large $c/a$ and polarizations we obtain at the MPB will stimulate
further experimental study of this promising system.\cite{fn:exp}

\section{Alloys: Piezoelectric response near the MPB}

Finally, we compute the piezoelectric response of BS-PT, and compare
with that of PZT. First we investigate the coefficients $e_{ij} =
dP_i/d\eta_j$ that describe the so-called {\sl inverse} piezoelectric
effect ($\eta_j$ is the $j$-th component of the strain tensor in Voigt
notation).\cite{fn:pro} We report here our results for $e_{15}$ of the
tetragonal 
phase, the coefficient known to be responsible for the large
piezoelectric response of PZT.~\cite{bel00b,wu02} We focus on VCA alloys
close to the MPB on the tetragonal side; specifically, we consider
BS-PT at $x=0.7$ (BS-PT$^{x=0.7}$) and PZT at $x=0.6$ (PZT$^{x=0.6}$).

Following Ref.~\onlinecite{bel99}, we split the piezoelectric
coefficient into two parts
\begin{equation}
e_{15}=e_{15,{\rm i}} +e_{15,{\rm c}} ,
\label{eq:e15tot}
\end{equation}
where $e_{15,{\rm c}}$ is computed at fixed internal atomic coordinates
(``clamped-ion'' term) and
\begin{equation}
e_{15,{\rm i}} = \sum_s \frac{ea}{v} Z^*_{11}(s) \frac{du_1(s)}{d\eta_5},
\label{eq:e15i}
\end{equation}
(``internal-strain'' term).  Here
$s$ runs over the atoms in the unit cell, $e$ is the
magnitude of the electron charge, $Z^*_{11}$ is the effective
charge associated with displacements $u_1$ along the $x$ direction,
$a$ is the lattice constant in the $x$ direction, and $v$ is the unit
cell volume.

\begin{figure}[b!]
\begin{center}
\includegraphics[width=2.5in]{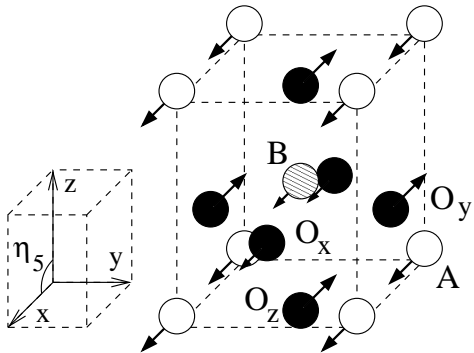}
\end{center}
\vskip 1mm
\caption{Sketch of the unit cell of the tetragonal phase of VCA BS-PT or
PZT. The arrows represent the displacements associated with $e_{15} =
dP_1/d\eta_5$. Atoms in the pairs A--O$_y$, A--O$_z$, and B--O$_x$
approach each other.}
\label{fig3}
\end{figure}

For BS-PT$^{x=0.7}$ we obtain $e_{15,{\rm c}}=-$0.93~C/m$^2$ and
$e_{15,{\rm i}}=8.18$~C/m$^2$, which add up to $e_{15}=7.25$~C/m$^2$.
For PZT$^{x=0.6}$ we obtain $e_{15,{\rm c}}=-$0.66~C/m$^2$,
$e_{15,{\rm i}}=$8.24~C/m$^2$, and $e_{15}=7.58$~C/m$^2$.  Hence, both
systems have a similar $e_{15}$, and in both the internal-strain
contribution strongly dominates the response. To understand this, we
note that a large value of $e_{15,{\rm i}}$ can arise in
Eq.~(\ref{eq:e15i}) either from a ``dielectric effect'' (large $Z^*$)
or an ``elastic effect'' (large response of the internal coordinates
to strain, i.e., large $du_1/d\eta_5$). Table~\ref{tab3} shows the
calculated values of $Z^*_{11}$ and $du_1/d\eta_5$ for all atoms in
the unit cell of VCA BS-PT$^{x=0.7}$ and PZT$^{x=0.6}$. In both
systems the response is mainly associated with an enhanced covalent
bonding between the A ion and oxygens O$_y$ and O$_z$, as pictorially
shown in Fig.~\ref{fig3}.
Our results further suggest that the substitution of Pb by Bi enhances
such a mechanism, consistent with the fact that Bi is more
electronegative than Pb. The B ion contributes by shifting closer to
O$_x$, but for the B--O$_x$ pair the elastic effect is substantially smaller
than it is in the case of A--O$_y$ and A--O$_z$. We also note that
when these results are compared with the analogous ones for the smaller
$e_{33}$ coefficient (not shown here -- see Refs.~\onlinecite{bel00a}
and \onlinecite{bel99} for related PZT results), one finds that it is
the elastic effect that makes $e_{15}$ particularly large.

The above results on $e_{15}$ have revealed much about the chemical
origin of the
piezoelectric response of these systems. However, the energetics of
the piezoelectric response, as well as the actual technological
interest of a material, is best characterized by the {\it direct}
piezoelectric coefficients $d_{ij} = dP_i/d\sigma_j$ reflecting the
polarization that develops in direction $i$ by application of a
{\it stress} with Voigt component $j$ (or equivalently, the strain
$j$ induced by an electric field along $i$).
In particular, $d_{15}$ measures the ease
of polarization rotation associated with the large responses at the
MPB. Technical difficulties have traditionally hampered the direct
calculation of $d_{ij}$ coefficients from first principles.\cite{fn:ds}
In the present case, however, it is not hard to show that $d_{15}$
is just given by $e_{15}/\widetilde{C}_5$, where $\widetilde{C}_5$ is the
shear elastic constant associated with $\eta_5$ when the atomic positions are
allowed to relax in response to the strain (i.e., the {\it
dressed} elastic constant) at fixed electric field. We calculated
$d_{15}=$168~pC/N (with $\widetilde{C}_5$=0.62~a.u.) for 
BS-PT$^{x=0.7}$ and $d_{15}=$196~pC/N (with $\widetilde{C}_5$=0.56~a.u.)
for PZT$^{x=0.6}$. These results 
agree in magnitude with the piezoelectric coefficients of ceramic
samples of BS-PT and PZT measured at room temperature and at the
MPB, as well as with the semiempirical calculation of Wu and Krakauer
for an ordered PZT supercell.\cite{wu02} 

\begin{table}
\caption{Effective charges $Z^*_{11}$ and derivatives $du_1/d\eta_5$
associated with the $e_{15,{\rm i}}$ piezoelectric coefficients of VCA
BS-PT$^{x=0.7}$ and PZT$^{x=0.6}$. Derivatives given in units of $a$, the
lattice parameter along the $x$ direction; $a$=7.27~a.u. for
BS-PT$^{x=0.7}$ and $a$=7.42~a.u. for PZT$^{x=0.6}$. We use the
convention that $\sum_s du_1(s)/d\eta_5=0$.}
\vskip 1mm
\begin{tabular}{ldddd}
  & \multicolumn{2}{c}{$Z^*_{11}$}
  & \multicolumn{2}{c}{$du_1/d\eta_5$} \\
  & \multicolumn{1}{c}{BS-PT$^{x=0.7}$} &
    \multicolumn{1}{c}{PZT$^{x=0.6}$}   &
    \multicolumn{1}{c}{BS-PT$^{x=0.7}$} &
    \multicolumn{1}{c}{PZT$^{x=0.6}$} \\
\hline
A     & 4.26 & 3.82 & 1.00 & 0.92 \\
B     & 5.27 & 6.64 & 0.07 & 0.16 \\
O$_x$ & $-$4.47 & $-$5.23 & 0.09 & 0.07 \\
O$_y$ & $-$2.77 & $-$2.90 & $-$0.41 & $-$0.51 \\
O$_z$ & $-$2.29 & $-$2.33 & $-$0.76 & $-$0.64
\end{tabular}
\label{tab3}
\end{table}

Thus, our calculations clearly provide evidence that BS-PT can be
expected to be a piezoelectric of comparable quality to PZT. This
conclusion, as well as the leading role of Bi and/or Pb in the
response, is further supported by calculations of $e_{33}$ in the
tetragonal phase of the above-mentioned 10-atom supercells (i.e., by
calculations in which the VCA was not used).

\section{Summary and conclusions}

Our first-principles study of BS-PT indicates that this material displays
very large structural distortions and polarizations at the MPB. In
particular, we obtain $c/a$ between 1.05 and 1.08 and $P_{\rm
tet}\approx 0.9$~C/m$^2$. We also find that the piezoelectric response
of BS-PT near the MPB is comparable to that of PZT. 

Our calculations show that the large polarization and piezoelectric
responses in BS-PT are mainly related with the onset of
Bi/Pb-6$p$--O-2$p$ hybridization, a mechanism that is enhanced upon
substitution of Pb by Bi, since Bi is a more covalent atom. Our
results also provide evidence of the dominant role of Pb in the
piezoelectric response of PZT, in agreement with recent experimental
studies.~\cite{ega02} In both BS-PT and PZT, this predominance of the
A-site ions is related to a large elastic effect, i.e., a large
displacement of the ions in response to strain.

Our results are in reasonably good agreement
with experiment. The quantitative discrepancies between theory and
experiment (in particular, for the values of $c/a$ and the
polarizations at the MPB) suggest that the intrinsic piezoelectric
properties of BS-PT alloys may be even better than those measured
experimentally to date.  We thus hope that our results will
stimulate further experimental work on this and related materials
systems.

\acknowledgments

J.~I. thanks J.~B.~Neaton for many stimulating discussions. This work
was supported by ONR Grant No. N0014-97-1-0048 and by the Center for
Piezoelectrics by Design (CPD) under ONR Grant N00014-01-1-0365.
Computational facilities for the work were also provided by the CPD.


\begin{thebibliography}{99}

\bibitem{uch96} K. Uchino, {\sl Piezoelectric Actuators and Ultrasonic
Motors} (Kluwer Academic Publishers, Boston, 1996).

\bibitem{par02} S.-E. Park and W. Hackenberger, Curr. Opin. Solid
State Mater. Sci. {\bf 6}, 11 (2002).

\bibitem{noh02} B. Noheda, Curr. Opin. Solid State Mater. Sci. {\bf
6}, 27 (2002).

\bibitem{eitel} R. E. Eitel, C. A. Randall, T. R. Shrout,
P. W. Rehrig, W. Hackenberger, and S.-E. Park,
Jap. J. Appl. Phys. {\bf 40}, 5999 (2001); R. E. Eitel, C. A. Randall,
T. R. Shrout, and S.-E. Park, Jap. J. Appl. Phys. {\bf 41}, 1 (2002).

\bibitem{coh92} R. E. Cohen, Nature {\bf 358}, 136 (1992).

\bibitem{pos94} M. Posternak, R. Resta, and A. Baldereschi,
Phys. Rev. B {\bf 50}, 8911 (1994).

\bibitem{ega02} T. Egami {\it et al.}, Proceedings of ``Fundamental
Physics of Ferroelectrics'' (Washington DC, 2002), R. E. Cohen,
ed. (AIP, Melville, New York, 2002), p. 216.

\bibitem{van90} D. Vanderbilt, Phys. Rev. B {\bf 41}, 7892 (1990).

\bibitem{kin93} R. D. King-Smith and D. Vanderbilt, Phys. Rev. B {\bf
47}, 1651 (1993).

\bibitem{bel00a} L. Bellaiche and D. Vanderbilt, Phys. Rev. B {\bf
61}, 7877 (2000).

\bibitem{ram00} N. J. Ramer and A. M. Rappe, Phys. Rev. B {\bf 62},
743 (2000).

\bibitem{fn:pureBS} This study could not be accomplished
experimentally, since perovskite BS is not stable at ambient
pressure.\protect\cite{eitel}

\bibitem{fn:quantum} The atomic relaxations of some of the systems
studied here are so large that very large $P$'s, approaching the
``quantum of polarization'',\protect\cite{kin93} might be
expected.  We thus made sure to remain on the correct polarization
branch by requiring a smooth behavior of $P(x)$.  However, the
reported polarizations turned out to remain small enough that this
problem did not really arise (e.g., compare with
$2e/a^2\simeq2.3$~C/m$^2$ for tetragonal BS).

\bibitem{kin94} See band structures for a family of ferroelectric
perovskites in Fig.~4 of R. D. King-Smith and D. Vanderbilt,
Phys. Rev. B {\bf 49}, 5828 (1994). Note that PZ is the only material
with a band structure similar to that of BS.

\bibitem{hil99} A significant Bi-6$p$--O-2$p$ hybridization was also
found in the magnetic perovskite BiMnO$_3$. See N.A. Hill and K.M. Rabe,
Phys. Rev. B {\bf 49}, 8759 (1999).

\bibitem{sha76} The ionic radii used in this article were taken from
R. D. Shannon, Acta Cryst. A {\bf 32}, 751 (1976).

\bibitem{for01} M. Fornari and D. J. Singh, Phys. Rev. B {\bf 63},
092101 (2001).

\bibitem{ini01} J. \'I\~niguez and L. Bellaiche, Phys. Rev. Lett. {\bf
87}, 095503 (2001).

\bibitem{geo01} A. M. George, J. \'I\~niguez, and L. Bellaiche, Nature
{\bf 413}, 54 (2001).

\bibitem{fu00} H. Fu and R. E. Cohen, Nature {\bf 403}, 281 (2000).

\bibitem{bel00b} L. Bellaiche, A. Garc\'{\i}a, and D. Vanderbilt,
Phys. Rev. Lett. {\bf 84}, 5427 (2000).

\bibitem{capzt} B. Noheda, D. E. Cox, G. Shirane, J. A. Gonzalo,
L. E. Cross, and S.-E. Park, Appl. Phys. Lett. {\bf 74}, 2059
(1999).

\bibitem{ber59} D. Berlincourt and H. H. A. Krueger,
J. Appl. Phys. {\bf 30}, 1804 (1959).

\bibitem{fn:exp} Ref.~\protect\onlinecite{eitel} reports that, as the
PT limit is approached from the MPB side, $P_{\rm tet}$ decreases as
$c/a$ increases, which is a very unusual behavior ($P_{\rm tet}$ is
found to grow with $c/a$ in all systems we are aware of). Also,
$P_{\rm tet}$ should reach a value of approximately 0.75~C/m$^2$ in
the PT limit,\cite{lin77} which seems inconsistent with the results of
Ref.~\onlinecite{eitel}.

\bibitem{lin77} M. E. Lines and A. M. Glass, {\it Principles and
Applications of Ferroelectrics and Related Materials} (Clarendon
Press, Oxford, 1977).

\bibitem{fn:pro} Here we discuss the {\it proper} piezoelectric
response, which we calculate as in D. Vanderbilt,
J. Phys. Chem. Sol. {\bf 61}, 147 (2000).

\bibitem{wu02} Z. Wu and H. Krakauer, Proceedings of ``Fundamental
Physics of Ferroelectrics'' (Washington DC, 2002), R. E. Cohen,
ed. (AIP, Melville, New York, 2002), p. 9.

\bibitem{bel99} L. Bellaiche and D. Vanderbilt, Phys. Rev. Lett.
{\bf 83}, 1347 (1999).

\bibitem{fn:ds} The only direct first-principles calculation of
$d_{ij}$ coefficients that we are aware of is the recent one of F.
Bernardini and V. Fiorentini [Appl. Phys. Lett. {\bf 80}, 4145
(2002)] who used a damped Parrinello-Rahman dynamics to study III-V
nitrides under applied stress.

\end{thebibliography}
\end{document}